# Short range order and topology of Te-rich amorphous Ge-Sb-Te alloys

I. Pethes[a,*], A. Piarristeguy[b], A. Pradel[b], R. Escalier[b], M. Micoulaut[c], S. Michalik[d], J. Darpentigny[e], A. Zitolo[f], P. Jóvári[a]

[a]HUN-REN Wigner Research Centre for Physics, Institute for Solid State Physics and Optics, Konkoly Thege út 29–33., H-1121 Budapest, Hungary

[b]ICGM, Univ Montpellier, CNRS, ENSCM, Montpellier, France

[c]Sorbonne Université, Laboratoire de Physique Théorique de la Matière Condensée, CNRS UMR 7600, Boite 121, 4 place Jussieu 75252 Paris Cedex 05, France

[d]Diamond Light Source Harwell Science and Innovation Campus, Didcot, Oxfordshire, OX11 0DE, UK

[e]Laboratoire Léon Brillouin, CEA-Saclay 91191, Gif sur Yvette Cedex, France

[f]Synchrotron SOLEIL, L'Orme des Merisiers, Départementale 128, 91190, France

Abstract

The structure of evaporated amorphous $Ge_xSb_xTe_{100-2x}$ ($x$ = 6, 9, 13) alloys was investigated by neutron diffraction, X-ray diffraction and extended X-ray absorption spectroscopy (EXAFS) at the Ge, Sb and Te K-edges. Large scale structural models were generated by fitting the experimental datasets (5 for each composition) simultaneously in the framework of the reverse Monte Carlo simulation technique. It was found that the alloys are chemically ordered (Ge and Sb have predominantly Te neighbors) and within the experimental uncertainty each component satisfies the 8 – N rule. A comparison with the pair correlation functions of melt quenched $Ge_{20}Te_{80}$ revealed that the first minimum of $g_{TeTe}(r)$ is shallower in the ternary alloys than in $Ge_{20}Te_{80}$. On the other hand, the separation of the first and second coordination environments of Ge atoms is stronger in the Ge-Sb-Te alloys investigated.

1. Introduction

Rapid and reversible crystallization of amorphous tellurides induced by electric field or light was reported more than 50 years ago [1, 2]. Since then the structure and physical properties of these materials have been intensely investigated.

---

* Corresponding author. E-mail address: pethes.ildiko@wigner.hun-ren.hu

In the first two decades, systematic structural investigations focused mostly on binary Ge-Te alloys. In 1970, Betts, Bienenstock and Ovshinsky studied $Ge_{11}Te_{89}$ and $Ge_{46}Te_{54}$ with X-ray diffraction [3]. They found that i) the short range order is different in crystalline and amorphous Ge-Te alloys and ii) the average Ge and Te coordination numbers are four and two, respectively. Nicotera et al. arrived to similar conclusions in their pioneering neutron diffraction study on $Ge_{17.5}Te_{82.5}$ [4]. Based on an analogy with amorphous black P, Bienenstock proposed a threefold coordinated structural model for amorphous GeS, GeSe and GeTe [5]. His model was supported by the neutron diffraction study of Pickart et al. [6]. The reason of the disparate conclusions was obviously the information deficient nature of the experimental data behind the models. Nearest neighbor Ge – Te and Te – Te distances are close to each other that leads to a strong correlation between the short range order parameters. This is well illustrated by the neutron diffraction study of Ichikawa et al. [7] who claimed the lack of Te – Te bonds and a 6.3 ± 0.4 average coordination number of Ge in $Ge_{20}Te_{80}$. In later experimental studies the situation was improved by the combination of X-ray diffraction, neutron diffraction and EXAFS datasets. Though structural parameters are still scattered to some extent the '4 – 2' model (Ge and Te coordination numbers are four and two, respectively) seems to be supported by most recent investigations [8].

This result is consistent with Mott's rule [9], which states that the total coordination number of an atom is $8 - N$, where $N$ is the number of valence electrons of the atom. Though there are some exceptions (e.g., Ge-Sb-S glasses [10]), this rule has been verified for several amorphous chalcogenide alloys composed of elements of groups 14-15-16 of the periodic table, see, e.g., [11-14].

Over the years, various models have been developed for the bonding preferences of the constituents. The most popular ones are the topologically ordered network model (TONM) [15-17] and the chemically ordered network model (CONM) [18]. In the TONM, all bond types are equally represented, and the physical properties of the glass are determined by the mean coordination number (the concentration-weighted sum of the coordination numbers of the elements). In the CONM, bonds between chalcogen (Ch) and non-chalcogen (nonCh) atoms are preferred, while the number of Ch – Ch and nonCh – nonCh bonds is minimized. This means that in a stoichiometric composition, only Ch – nonCh bonds are present, while Ch – Ch bonds can be found in Ch-rich glasses and nonCh – nonCh bonds in Ch-deficient alloys.

Chemical ordering is most prominent in sulfide glasses, as observed in various studies (e.g., [10, 19-21]). While bonding preferences were also observed in selenide and telluride glasses, some 'wrong bonds' were reported in these systems (e.g., [8, 13, 14, 22-24]). These bonding defects become especially visible by neutron diffraction with isotopic substitution performed on selected chalcogenide systems [25], and independently confirmed by *ab initio* molecular dynamics simulations [26, 27].

In amorphous $Ge_xTe_{100-x}$ alloys, where $N_{Ge} = 2$ and $N_{Te} = 4$ (see above), the composition $Ge_{33}Te_{67}$ is stoichiometric, meaning that only Ge – Te bonds should be present according to the CONM. Te-poor alloys ($x > 33$) are expected to have Ge – Ge bonds, while over-stoichiometric glasses (Te-rich, $x < 33$) should have Te – Te bonds [28]. It has been discovered that when the Ge content is less than 20%, the presence of Ge – Ge bonds is insignificant (chemically ordered glass). However, when the Ge content is equal to or greater than 24%, Ge – Ge bonds have been observed. It is important to note that the number of these bonds is significantly lower than what would be present in a random covalent glass [8]. In Ge-As-Te glasses, As – As bonds have been identified in Te-rich compositions, while Te – Te bonds were present in Te-deficient compositions [14].

'Phase-change' Ge-Sb-Te alloys like $Ge_2Sb_2Te_5$ (GST-225) and $Ge_1Sb_2Te_4$ (GST-124) are Te-deficient. In these compositions the number of Te – Te bonds is negligibly low, and the presence of Ge/Sb – Ge/Sb bonds is significant, consistently with the CONM model (see [12, 29]). The Ge-Sb-Te system has a very small glass forming region (see Fig. 1 of Ref. [30]). Before 2021, stable glasses prepared by melt quenching technique were only reported in the vicinity of $Ge_{15}Te_{85}$ and $Ge_{20}Te_{80}$ binary compositions with ≤ 5% antimony content ([31-33]). As far as we know, the only reported structural studies of Te-rich Ge-Sb-Te glasses were the papers of Saiter et al. [34,35], in which $Ge_{15}Sb_xTe_{85-x}$ glasses with 0.5%, 3% and 5% Sb content were investigated by EXAFS measurements at the Ge K absorption edge. The authors found that in these Te-rich glasses, the Ge atoms are surrounded by 4 Te atoms, and Ge-Ge or Ge-Sb bonds are not present, in agreement with the CONM.

Piarristeguy et al. have recently produced amorphous thick films along the $Ge_xSb_xTe_{100-2x}$ tie-line [30], even at higher antimony content. While first principle molecular dynamics simulations were reported about such alloys previously [36] including the liquid phase [23], experimental structural data were missing until now. Here we present the first comprehensive experimental study on the structure of Te-rich amorphous Ge-Sb-Te alloys. The measurements include neutron and X-ray diffraction, as well as EXAFS data at the Ge, Sb, and Te absorption edges. These datasets are analyzed using the reverse Monte Carlo (RMC) simulation method.

Experimental structure determination of ternary amorphous alloys is not an easy task. In case of the Ge-Sb-Te system the situation is complicated by the similar size and scattering power of Sb and Te. Therefore, the simulation procedure will be discussed in detail below. The information we can obtain from the experimental data-driven models will also be analyzed thoroughly.

As this work reports the results of a reverse Monte Carlo simulation study of Te-rich amorphous Ge-Sb-Te alloys it may be useful to summarize briefly the limitations and strengths of this technique. Unless energetic or geometrical constraints are used RMC simulations are driven by the experimental data. Poor data quality, short data range or insufficient number of datasets may all result in unreliable models. Real space resolution is always finite due to the sampling (Nyquist) theorem [37], and the separation of partial pair correlation functions can be impossible if the number of components is too high. The last fundamental limitation is that diffraction and EXAFS datasets depend directly on pair correlations thus higher order correlations (e.g. bond angle distributions) may not be pinned down even if partial pair correlation functions can be separated. These features should always be kept in mind when models generated by RMC are discussed.

We would also like to mention two positive traits of this method. The first one is its flexibility: a large number of data sets (including neutron, electron and X-ray diffraction, EXAFS and anomalous scattering) and a large variety of constraints (average coordination number, coordination number distribution, second neighbor distribution, bond angle) can be used to build up structural models. The second is that *in favorable cases* (high data quality, sufficient number of data sets…) the error of coordination numbers can be as low as a few percent and the uncertainty of nearest neighbor distances is about 0.02 – 0.03 Å. It should be emphasized that if constraints are not applied then RMC uses only the most basic physical knowledge of the system investigated (density, minimum interatomic distances), thus – keeping in mind the above limitations – it can be considered as a simple and robust technique for combining and inverting diffraction and EXAFS datasets to a set of partial pair correlation functions.

2. Experimental

2.1. Sample preparation

Three Ge-Sb-Te films (thicknesses 6-7 μm) of nominal compositions $Ge_6Sb_6Te_{88}$, $Ge_9Sb_9Te_{82}$, and $Ge_{13}Sb_{13}Te_{74}$ were deposited on glass substrates by thermal co-evaporation of the pure elements (Ge (Goodfellow, lump, 99.999%), Sb (Sigma-Aldrich, beads, 99.999%), and Te (Sigma-Aldrich, pieces, 99.999%) using a PLASSYS MEB 500 device equipped with two current induced heating sources and an electron beam evaporator. The three sources were placed in a configuration that allowed the deposition of films with uniform composition and thickness over a surface of about 4 cm in diameter. The germanium was evaporated using an electron beam, while the two current-induced heated sources were utilized for evaporating antimony and tellurium. Each material was housed in specific crucibles to maintain stable evaporation rates [30]. Glass substrates were cleaned with alcohol and dried with dry air. Prior to deposition, the chamber was evacuated to approximately $10^{-5}$ Pa. During the deposition process, the substrate holder rotated at 8 rpm. The evaporation rate (~ 420 nm/min) and thickness for each element were automatically controlled using pre-calibrated quartz crystal monitors. No further annealing treatment was conducted before proceeding to film characterization. Powders used for structural measurements were obtained by scraping films from the substrate. The amorphous nature was checked by X-ray diffraction using a PANalytical XPERT diffractometer. A Cu Kα source (λ = 1.5406 Å) was used for the excitation with operating voltage of 40 kV and a beam current of 30 – 40 mA.

2.2. Measurements

The neutron diffraction experiment was carried out at the 7C2 liquid and amorphous diffractometer [38] installed at the Orphée research reactor (Saclay, France). The instrument was equipped with a detector system consisting of 256 position sensitive tubes containing $^3$He. Samples were filled into thin walled vanadium sample holders of 5 mm diameter. Wavelength and detector position were determined by measuring Ni powder. The wavelength of incident neutrons was 0.724 Å. Vanadium powder was also measured to take into account detector efficiency. Raw data were corrected for background, multiple and incoherent scattering, and absorption using standard procedures.

The high energy X-ray diffraction measurements were performed on beamline I15-1 [39] at Diamond Light Source (Didcot, UK). The X-ray beam energy was 76.69 keV (corresponding to the wavelength of 0.161669 Å). The powder sample material was loaded into a capillary with 1.2 mm diameter. The capillary was illuminated by an X-ray beam of 0.7 x 0.15 mm$^2$ for 300 seconds. The scattered signals were recorded in transmission mode by a flat panel detector (Perkin Elmer XRD 4343 CT) at a sample-to-detector distance of 196.8 mm. Two dimensional diffraction patterns were azimuthally integrated to obtain intensity versus $Q$ curves employing the software DAWN [40]. The precise energy and experimental geometry (e.g., beam centre position, detector orthogonality) was obtained by fitting a CeO$_2$ NIST standard data collected at multiple sample-to-detector distances [41].

Raw intensity $I(Q)$ data were corrected for the background contribution (air and empty capillary), self-absorption, fluorescence, Compton scattering and normalized to the electron unit by the PDFgetX2 software [42]. Normalized elastic scattering intensities were converted to structure factors $S(Q)$ using the Faber-Ziman equation [43].

Corrected neutron- and X-ray diffraction structure factors are shown in Figs. 1 and 2.

Ge, Sb and Te K-edge EXAFS data were collected in transmission mode at the SAMBA station of Soleil (Saclay, France). Radiation from a bending magnet source was monochromatized by a sagittally focusing Si 220 monochromator. Finely ground samples were mixed with cellulose and pressed into disks of 10 mm diameter and ~ 1 mm thickness. Intensities before and after the samples were measured by ionization chambers (IC0 and IC1, respectively). For the Ge K-edge measurement a mixture of He (400 mbar) and N$_2$ (600 mbar) was used for IC0 while IC1 was filled with Ar (100 mbar) and N$_2$ (900 mbar). For the Sb and Te K-edge scans a mixture of Ar (500 mbar) and $N_2$ (500 mbar) was applied in IC0 and IC1 was filled with 1000 mbar of Ar.

Ge-, Sb-, and Te K-edge $\chi(k)$ curves were obtained by standard procedures of data reduction using the program VIPER [44]. Raw $k^3\chi(k)$ data between 2 Å$^{-1}$ and 14-15 Å$^{-1}$ were Fourier-transformed into $r$-space using a Kaiser-Bessel window (α = 1.5). The as obtained real space data were then multiplied by a rectangular window (the $r$-space ranges were 1.6 Å – 2.9 Å, 1.8 Å – 3.1 Å and 1.5 Å – 3.1 Å for Ge, Sb, and Te EXAFS data, respectively) and transformed back to $k$-space. The resulting $\chi(k)$ functions were used in the reverse Monte Carlo simulations. Backscattering factors needed for the calculation of model $\chi(k)$ functions [45] were obtained by the program feff8.4 [46].

3. Reverse Monte Carlo simulations

The reverse Monte Carlo method [47] is a robust tool to get large three-dimensional structural models that are consistent with experimental data, in particular the total structure factors obtained from neutron diffraction (ND) or X-ray diffraction (XRD) experiments, and extended X-ray absorption fine structure (EXAFS) curves. The simulation minimizes the discrepancies between experimental and model curves by randomly moving the particles and generates particle configurations compatible with all experimental data sets within their experimental errors. From the obtained configurations short range order parameters (partial pair correlation functions, average coordination numbers, nearest neighbor distances, angle distributions, etc.) can be calculated. Short range order parameters (coordination numbers and bond angle distributions, common neighbors of two atoms, second neighbors) can also be fixed by constraints to see whether certain values are compatible with available experimental data. The constraints used in this work are discussed in detail below.

In the present study, neutron and X-ray diffraction structure factors and Ge, Sb, and Te K-edge EXAFS datasets were fitted using the RMC++ code [48].

The investigated samples and their estimated number densities are collected in Table 1. The densities were determined by extrapolating the molar volumes of amorphous $Ge_{15}Te_{85}$ [49] and $Ge_{50}Te_{50}$ [50], assuming that the partial molar volumes of Sb and Te are equal (due to the low fraction of Sb, the error arising from this assumption is probably not significant). The cubic simulation boxes contained 10 000 atoms in test runs and 30 000 particles in the final runs presented here. Initial configurations were obtained by randomly placing the atoms in the simulation box and moving them around until their separations became higher than the minimum distances between atoms (cutoffs). Starting values of the cutoffs were usually around 90% of the sum of the corresponding atomic radii [51]; the final values are shown in Table 2. All the samples studied are Te-rich (over-stoichiometric), so the presence of Ge – Te, Sb – Te, and Te – Te pairs was allowed in all simulation runs. The necessity of Ge – Ge, Ge – Sb, and Sb – Sb pairs was investigated for the lowest Te-content $Ge_{13}Sb_{13}Te_{74}$ sample. It was found that the quality of the fits was not improved when these bonds were allowed. In the final models, Ge – Ge, Ge – Sb, and Sb – Sb bonds were thus forbidden using higher cutoff values than the expected bond lengths.

Some 'background' coordination constraints were always used to avoid isolated atoms or unphysically low coordination numbers (0 for Te, 0 and 1 for Sb, and 0, 1, and 2 for Ge). In the so-called 'unconstrained' models only the above coordination constraints were used.

To evaluate the validity of the Mott-rule in these glasses (and subsequently, reduce the degree of freedom of the fitting procedure) coordination constraints were applied: all Ge and Sb atoms were forced to have 4 and 3 neighbors, respectively. About 95% of the atoms met these requirements.

The different test models were classified according to their 'goodness-of-fit' (R-factor) values:

$$R = \frac{\sqrt{\Sigma_i \left(S_{\text{mod}}(Q_i) - S_{\text{exp}}(Q_i)\right)^2}}{\sqrt{\Sigma_i S_{\text{exp}}^2(Q_i)}} \qquad (1)$$

(Here, 'mod' denotes the model and 'exp' the experimental curves, while $Q_i$ are the experimental points. A similar expression is valid for the EXAFS curves.)

The average (partial) coordination numbers ($N_{ij}$) were calculated by integrating the partial pair correlation functions up to the first minimum ($r_{\min}$):

$$N_{ij} = 4\pi\rho_0 c_j \int_0^{r_{\min}} r^2 g_{ij}(r) \mathrm{d}r \qquad (2)$$

where $\rho_0$ is the density, and $c_j$ is the concentration of the *j*th element. (since $g_{ij}(r) = g_{ji}(r)$, therefore $N_{ji} = N_{ij}c_i/c_j$) Dedicated simulation runs were performed to estimate the uncertainty of the mean coordination numbers. The value of the tested $N_{ij}$ was systematically changed (± 5% steps). The range of $N_{ij}$ values, in which the quality of the fit is adequate can be determined by monitoring the *R*-factors.

4. Results and discussion

   4.1. Short range order parameters

Experimental total structure factors ($S(Q)$) and filtered, $k^3$-weighted EXAFS curves ($k^3\chi(k)$) are shown in Figs. 1 – 5.

The average coordination numbers of Ge and Sb obtained in the unconstrained model (in which only Ge – Te, Sb – Te, and Te – Te pairs were present) mostly obey the Mott-rule for each composition, ranging from 3.7 to 4.0 for Ge and around 2.75 for Sb (2.72 – 2.79). In the final model, Ge and Sb atoms were forced to have 4 and 3 neighbors, respectively, to reduce the uncertainties. The quality of the fit of this final model was as good as those of the unconstrained models. Model curves are compared with experimental ND, XRD, and EXAFS data in Figs. 1 – 5.

Partial pair correlation functions obtained from the final model are shown in Fig. 6. The average coordination numbers and nearest neighbor distances are given in Tables 3 and 4.

The total coordination number of Te atoms ($N_{Te} = N_{TeGe} + N_{TeSb} + N_{TeTe}$) was freely variable, not only in the unconstrained model but also in the final model where the coordination numbers $N_{TeGe}$ and $N_{TeSb}$ were fixed by the coordination constraints applied, but the value of $N_{TeTe}$ remained free. The total coordination number $N_{Te}$ is remarkably close to 2 in both models for all compositions: its value is 1.97 – 1.98 in the final models and 2.01 – 2.03 in the unconstrained models.

The uncertainty of the average Te – Te coordination number ($N_{TeTe}$) was determined by dedicated simulation runs in which the $N_{GeTe}$ and $N_{SbTe}$ average coordination numbers were fixed at 4 and 3, respectively, and the value of $N_{TeTe}$ was systematically varied. Similar tests were carried out to estimate the uncertainties of $N_{GeTe}$ and $N_{SbTe}$ average coordination numbers (only one of the above coordination numbers was changed at a time, the other was kept at its 'Mott-value'). The quality of the fits was compared based on their $R$-factors. The estimated uncertainties are given in Table 3.

From these simulation runs, it can be concluded that in the amorphous Te-rich Ge-Sb-Te alloys investigated all components obey the Mott-rule.

Some previous experimental and simulation results on the total coordination number of Ge, Sb, and Te atoms in binary Ge-Te and Sb-Te, and the ternary Ge-Sb-Te systems are collected in Table 5. Experimental results support the Mott-rule for Ge (3.7 – 4.24) and Sb (2.8 – 3.22). The values are more scattered for Te atoms (1.5 – 2.7), being closer to 2 in studies combining several experimental techniques. The results obtained in theoretical (*ab initio* molecular dynamics, AIMD) studies depend strongly on the applied exchange-correlation functional and pseudopotential, on the consideration of long range van der Waals dispersion forces (DFT-D2), and even on the size of the simulation box (see, e.g., [60,77,79]). Of special interest is the inclusion of van der Waals correction forces and the DFT-D2 scheme which lead to an improved agreement between the calculated Ge – Te bond distances and the experimental values [23,24]. Earlier *ab initio* simulations tend to give a longer Ge – Te bond length [58,70] and overestimate the octahedral environment of Ge [80,81].

Nearest neighbor distances obtained in the final model are presented in Table 4, while some of the previous experimental and simulation results are collected in Table 5. The $r_{GeTe}$ value (2.60 Å) agrees well with experimental results for binary amorphous $Ge_xTe_{100-x}$ and Te-poor Ge-Sb-Te compounds. Recent AIMD simulations using the Becke-Lee-Yang-Parr (BLYP) exchange-correlation functional and the Troullier-Martins (TM) pseudopotential have resulted in similar Ge – Te distances: 2.59 – 2.62 Å in Ge$_x$Te$_{100-x}$ [60], and 2.63 Å in Ge$_2$Sb$_2$Te$_5$ [77, 79].

The same level of agreement was achieved by using different exchange-correlation functionals, but taking into account the long-range van der Waals forces in Ge$_x$Te$_{100-x}$ [59, 60] and in Te-rich Ge-Sb-Te glasses [36].

The $r_{SbTe}$ distance (2.83 – 2.84 Å) agrees well with the value found previously in Pd-Ge-Sb-Te (2.84 Å) [54] and Sb-Te films (2.83 – 2.86 Å) [61], as well as in amorphous Ge$_2$Sb$_2$Te$_5$ and GeSb$_2$Te$_4$ (2.82 – 2.85 Å) [12, 29, 64, 65, 67]. Theoretical studies give Sb-Te bond lengths mostly between 2.88 and 3.0 Å. Exceptions are the work of Hegedüs and Elliott [70] that reported 2.82 Å, and that of Kim et al. [75], with values between 2.79 and 2.93 Å, depending again on the details of the simulation, and especially the effect of the dispersion van der Waals correction.

The $r_{TeTe}$ bond length (2.76 – 2.77 Å) is in good agreement with most previously reported experimental values in amorphous Ge-Te (2.73 – 2.78 Å [8, 53, 56]), but shorter than that obtained from Te EXAFS data alone (2.8 – 2.82 Å [55]). The results of AIMD simulations are scattered in the 2.83 – 2.95 Å range. Simulations including the long range van der Waals dispersion forces resulted in shorter Te – Te bond distances [60].

4.2. First minima of the Te – Te, Ge – Te, and Sb – Te partial pair correlation functions

Fig. 7 shows $g_{TeTe}(r)$ of the amorphous Ge-Sb-Te alloys investigated, together with the $g_{TeTe}(r)$ function of melt quenched Ge$_{20}$Te$_{80}$ [8]. In amorphous Ge-Sb-Te, at $r \approx 3.0 - 3.1$ Å $g_{TeTe}(r)$ increases with increasing Sb content, indicating that the first and second coordination shells of Te overlap. Dedicated simulation runs revealed that Te – Te pairs with ≈ 3 Å separation cannot be eliminated without degrading the quality of the fits or introducing artificially sharp features in $g_{TeTe}(r)$. It implies that such Te – Te distances must be present in the amorphous Ge-Sb-Te alloys investigated.

The first minimum region of $g_{GeTe}(r)$ functions is also different in Ge$_{20}$Te$_{80}$ and Ge-Sb-Te alloys. The Ge – Te partial pair correlation function of melt quenched Ge$_{20}$Te$_{80}$ starts to rise at 3.2 Å, while all three $g_{GeTe}(r)$ functions of the Ge$_x$Sb$_x$Te$_{100-2x}$ alloys vanish between 3 Å and 3.5 Å (see Fig. 8). In case of sputtered amorphous Ge$_2$Sb$_2$Te$_5$ [12] the minimum is shallow, indicating the existence of Ge – Te distances between 2.8 Å and 3.3 Å (we note that the small maximum at 3.2 Å is an artefact connected to the upper limit of EXAFS fitting range). The pronounced shoulder of $g_{GeTe}(r)$ of Ge$_2$Sb$_2$Te$_5$ shows that a part of Ge atoms is in octahedral environment [71].

Though to a lesser extent, the minimum of the Sb – Te partial pair correlation functions also becomes deeper in the 3.1 Å – 3.4 Å range with decreasing Sb content (Fig. 9). The origin of the above differences of $g_{TeTe}(r)$, $g_{GeTe}(r)$, and $g_{SbTe}(r)$ functions may be clarified by further investigations (e.g., by a systematic comparison of the structure of sputtered and evaporated Ge-Sb-Te alloys).

### 4.3. Second neighbor Ge – Ge pairs

For all compositions $g_{GeGe}(r)$ curves obtained by unconstrained simulations show a double peak in the 3.5 – 5.5 Å region. Previously, in binary $Ge_{18.7}Te_{81.3}$, a sharp peak of the $g_{GeGe}(r)$ was found at $r \approx 3.8$ Å [8], which originates mainly from the corner-sharing Ge-centered tetrahedra. Several dedicated simulations were carried out on the $Ge_{13}Sb_{13}Te_{74}$ composition (which has the highest Ge content and thus the highest weight of Ge – Ge pairs in the experimental data) to investigate the second neighbor Ge – Ge pairs in the ternary Ge-Sb-Te alloys.

First, a simulation was performed in which only the XRD and 3 EXAFS data sets were fitted, without the neutron diffraction data. The shape of the $g_{GeGe}(r)$ curve of $Ge_{13}Sb_{13}Te_{74}$ with and without ND data is shown in Fig. 10. The impact of the ND data is twofold: (1) there are no Ge – Ge (second neighbor) pairs in the 3 – 3.5 Å region, (2) $g_{GeGe}(r)$ strongly increases between 3.5 Å and 4.5 Å. (We note that in this composition the weight of the Ge – Ge partial is only 3% in the neutron diffraction structure factor, while it is below 1% in the X-ray data set. Even such a small difference may have an impact on the generated models.)

In a second simulation, all Ge and Sb atoms had to have 4 and 3 neighbors, respectively, the Te – Ge – Te bond angle was constrained to be in the 109.5° ± 15° range and a constraint was also used to center the Ge – Te – Ge bond angle distribution at 101°. As a result, the second Ge – Ge peak in the 4.5 – 5.5 Å range was eliminated (see Fig. 10). The quality of the fit of the experimental data sets was as good as without these angle constraints therefore it is reasonable to assume that – similarly to binary Ge-Te glasses – Te-rich amorphous Ge-Sb-Te alloys also have a well-defined Ge – Ge peak.

Configurations obtained by the above set of constraints were further analyzed: Ge centered tetrahedra sharing one or two common Te neighbors (corner-shared, CS; or edge-shared ES) were determined. It was found that the dominant contribution to the first peak of the $g_{GeGe}(r)$ comes from corner sharing Ge-centered tetrahedra, see Fig. 11. Edge-sharing tetrahedra are observed at Ge – Ge distances around 3.6 Å, while topologically distant Ge – Ge pairs are dominant in the $r > 4.5$ Å region.

### 4.4. Comparison with some other chalcogenide glasses

The validity of the Mott-rule and the observation that the experimental data sets can be fitted with the model in which only Ge – Te, Sb – Te, and Te – Te pairs are present means that the structure of these amorphous alloys can be described with the CONM. This result is comparable with those obtained on the analogous Ge-Sb-Ch (Ch = S, Sb, Te) systems: Ge-Sb-S glasses [10], Ge-Sb-Se glasses [13] and Te-poor amorphous Ge-Sb-Te alloys [12]. The experimental techniques and simulation methodology applied in the four investigations are essentially the same: diffraction (XRD and ND) and EXAFS data sets were fitted simultaneously using the RMC technique. In all systems, the majority of atoms were found to follow the Mott-rule. The structure of these systems is best described by the CONM: the most prominent bonds are the Ge – Ch and Sb – Ch bonds. Ch – Ch bonds are found only in Ch-rich compositions. In Ch-poor systems, Ge – Ge and/or Ge – Sb pairs are also present.

5. Conclusions

Te-rich Ge-Sb-Te alloys were investigated by neutron and X-ray diffraction techniques and EXAFS. Models compatible with available experimental evidence revealed that the alloys are chemically ordered and all components satisfy the $8 - N$ rule. Neighboring GeTe$_4$ tetrahedra have in most cases one common Te atom (corner sharing). The minimum of $g_{TeTe}(r)$ at about 3 Å is not as deep as in melt quenched Ge$_{20}$Te$_{80}$ showing that the separation of the first and second coordination shells of Te is stronger in binary Ge-Te. The opposite was observed for the environment of Ge atoms: $g_{GeTe}(r)$ functions of the Ge$_x$Sb$_x$Te$_{100-2x}$ alloys studied are all zero between 3.0 Å and 3.5 Å.

Acknowledgment

I. P. and P. J. were supported by the ELKH (Eötvös Loránd Research Network, present name: HUN-REN) project 'Structure of materials used in energy storage' (Grant No. SA-89/2021). The neutron diffraction experiment was carried out at the ORPHÉE reactor, Laboratoire Léon Brillouin, CEA-Saclay, France.

**Table 1** Investigated Ge-Sb-Te glasses and their estimated number densities.

| Nominal composition | Number density [Å$^{-3}$] |
| --- | --- |
| $Ge_6Sb_6Te_{88}$ | 0.0271 |
| Ge$_9$Sb$_9$Te$_{82}$ | 0.0274 |
| Ge$_{13}$Sb$_{13}$Te$_{74}$ | 0.0279 |

**Table 2** Minimum interatomic distances used in the reverse Monte Carlo simulation [in Å]

| Bond type | Ge-Ge | Ge-Sb | Ge-Te | Sb-Sb | Sb-Te | Te-Te |
|---|---|---|---|---|---|---|
| Cutoff [Å] | 3.05 | 3.05 | 2.35 | 3.15 | 2.55 | 2.55 |

**Table 3** Coordination numbers of the investigated glasses obtained for the final model. The values highlighted in bold were constrained (see text). The values obtained by the so called 'unconstrained' model are also shown in square brackets along with the estimated uncertainties (see text for details).

| Pair (upper limit) | Ge$_6$Sb$_6$Te$_{88}$ | Ge$_9$Sb$_9$Te$_{82}$ | Ge$_{13}$Sb$_{13}$Te$_{74}$ |
|---|---|---|---|
| $N_{Ge-Te}$ (2.8 Å) | **4.01** [3.97 (-0.7 +0.8)] | **4.02** [3.96 (-0.3 +0.55)] | **3.99** [3.73 (-0.25 +0.45)] |
| $N_{Te-Ge}$ (2.8 Å) | **0.27** [0.27 (-0.05 +0.06)] | **0.44** [0.44 (-0.03 +0.05)] | **0.70** [0.66 (-0.05 +0.08)] |
| $N_{Sb-Te}$ (3.1 Å) | **2.99** [2.73 (-0.3 +0.9)] | **3.00** [2.79 (-0.1+1.1)] | **3.00** [2.72 (-0.2 +0.7)] |
| $N_{Te-Sb}$ (3.1 Å) | **0.20** [0.19 (-0.03 +0.06)] | **0.33** [0.31 (-0.01+0.12)] | **0.53** [0.48 (-0.03 +0.13)] |
| $N_{Te-Te}$ (3.1 Å) | 1.51 [1.56 (-0.19 +0.09)] | 1.2 [1.26 (-0.22 +0.08)] | 0.75 [0.89 (-0.36 +0.04)] |
| $N_{Te}$ | 1.98 [2.02 (±0.15)] | 1.97 [2.01 (±0.15)] | 1.98 [2.03 (±0.2)] |

**Table 4** Nearest neighbor distances [in Å]. The estimated uncertainty of the values is 0.02 Å.

| | Ge-Te | Sb-Te | Te-Te |
|---|---|---|---|
| Ge$_6$Sb$_6$Te$_{88}$ | 2.60 | 2.83 | 2.77 |
| Ge$_9$Sb$_9$Te$_{82}$ | 2.60 | 2.83 | 2.77 |
| Ge$_{13}$Sb$_{13}$Te$_{74}$ | 2.61 | 2.84 | 2.76 |

**Table 5** Total coordination numbers of Ge, Sb, and Te atoms and nearest neighbor distances [in Å] obtained by different experimental and simulation techniques in binary Ge-Te, and Sb-Te and ternary

Ge-Sb-Te systems. (Abbreviations: AXS: anomalous X-ray scattering; PBE: Perdew-Burke-Ernzerhof exchange-correlation functional; TM: Troullier-Martins pseudopotential; GTH: Goedecker-Teter-Hutter pseudopotential; TPSS: Tao-Perdew-Staroverov-Scuseria functional; PBEsol: modified PBE functional for solids; PAW: projected-augmented-wave pseudopotential; HSE: Heyd-Scuseria-Ernzerhof hybrid functional; vdW: van der Waals dispersion forces; BLYP: Becke-Lee-Yang-Parr exchange-correlation functional; ML-MD: machine-learning based molecular dynamics simulation. The number of atoms in the simulation boxes of AIMD calculations are given in parentheses.)

| Method | Compositions | $N_{Ge}$ | $N_{Sb}$ | $N_{Te}$ | $r_{Ge-Te}$ | $r_{Sb-Te}$ | $r_{Te-Te}$ | Ref. |
|---|---|---|---|---|---|---|---|---|
| EXAFS (Ge) | $Ge_{49}Te_{51}$ | 3.7 | | | 2.65 | | | [52] |
| ND | $Ge_{16}Te_{84}$, $Ge_{20}Te_{80}$ | 4 | | 2 | 2.59 | | 2.76 | [53] |
| EXAFS (Te) | $Ge_{52}Te_{48}$ | | | 1.5 | 2.59 | | | [54] |
| EXAFS (Te) | $Ge_{15}Te_{85}$, $Ge_{20}Te_{80}$ | | | 2 | 2.61 2.62 | | 2.8 2.82 | [55] |
| RMC; AXS | $Ge_{50}Te_{50}$ | 3.73 | | 2.52 | 2.6 | | 2.73 | [56] |
| EXAFS (Ge/Te) | $Ge_{47}Te_{53}$ | | | | 2.59/ 2.61 | | | [57] |
| RMC; ND, XRD, EXAFS (Ge) | $Ge_{14.5}Te_{85.5}$ $Ge_{18.7}Te_{81.3}$ $Ge_{23.6}Te_{76.4}$ | 4.16 4.10 4.06 | | 2.04 2.00 2.06 | 2.61 2.61 2.61 | | 2.75 2.76 2.75 | [8] |
| AIMD; PBE, TM (216) | $Ge_{50}Te_{50}$ | 4.2 | | 3.3 | 2.78 | | | [58] |
| AIMD; PBEsol+vdW (200) | $Ge_{20}Te_{80}$ | 4.17 | | 2.9 | 2.64 | | 2.9 | [59] |
| AIMD; PBE, TM (185) | $Ge_{20}Te_{80}$ | 4.54 | | 3.78 | 2.65 | | 2.88 | [60] |
| AIMD; PBE+vdW, TM (185) | $Ge_{20}Te_{80}$ | 4.35 | | 3.51 | 2.65 | | 2.83 | [60] |
| AIMD; BLYP, TM (215) | $Ge_{20}Te_{80}$ | 4.14 | | 2.57 | 2.62 | | 2.89 | [60] |
| AIMD; BLYP+vdW, TM (215) | $Ge_{20}Te_{80}$ | 3.97 | | 2.31 | 2.59 | | 2.84 | [60] |
| EXAFS (Sb/Te) | $Sb_{75}Te_{25}$ | | 2.8 | 2 | | 2.86/ 2.83 | | [61] |
| AIMD; PBE, GTH (240) | $Sb_2Te_3$ | | 4.09 | 2.74 | | 2.93 | 2.93 | [62] |
| AIMD; PBE, PAW (200) | $Sb_2Te_3$ | | 3.77 | 2.58 | | | | [63] |
| EXAFS (Ge) | $Ge_{15}Sb_xTe_{85-x}$ $x = 0.5, 3, 5$ | 4 | | | 2.60 | | | [34] |
| EXAFS (Te) | $Pd_1Ge_{17}Sb_{26}Te_{56}$ | | | 2 | 2.61 | 2.84[a] | 2.84[a] | [54] |
| EXAFS (Ge, Sb, Te) | $Ge_2Sb_2Te_5$ | | | | 2.61 | 2.85 | | [64] |

| Method | Composition | | | | | | | Ref. |
|---|---|---|---|---|---|---|---|---|
| EXAFS (Ge, Sb, Te) | Ge$_2$Sb$_2$Te$_5$ | 3.9 | 2.8 | 2.4 | 2.63 | 2.83 | | [65] |
| RMC; XRD | Ge$_2$Sb$_2$Te$_5$ | 3.7 | 3 | 2.7 | | | | [66] |
| RMC; XRD, ND, EXAFS (Ge, Sb, Te) | Ge$_2$Sb$_2$Te$_5$ | 4.24 | 3.22 | 2.04 | 2.64 | 2.83 | | [29] |
| RMC; XRD, ND, EXAFS Ge, Sb, Te) | Ge$_2$Sb$_2$Te$_5$<br>Ge$_1$Sb$_2$Te$_4$ | 3.85<br>3.91 | 3.12<br>2.91 | 1.99<br>1.98 | 2.60<br>2.61 | 2.82<br>2.83 | | [12] |
| RMC; AXS | Ge$_2$Sb$_2$Te$_5$ | 4.24 | 2.95 | 2.3 | 2.65 | 2.82 | | [67] |
| AIMD; PBE, TM (460) | Ge$_2$Sb$_2$Te$_5$ | 4.2 | 3.7 | 2.9 | 2.78 | 2.93 | | [58] |
| AIMD; PBE, GTH (270) | Ge$_2$Sb$_2$Te$_5$ | 3.82 | 4.03 | 2.87 | 2.79 | 2.94 | | [68] |
| AIMD; PBE, TM (460) | Ge$_2$Sb$_2$Te$_5$ | | | | 2.78 | 2.93 | 2.95 | [69] |
| AIMD; PBE, TM (63-90) | Ge$_2$Sb$_2$Te$_5$ | | | | 2.7 | 2.82 | | [70] |
| AIMD+RMC; TPSS, XRD (460) | Ge$_2$Sb$_2$Te$_5$ | 3.9 | 3.4 | 2.6 | 2.75 (2.65[b]) | 2.85 | | [71] |
| AIMD; PBEsol, TM (630) | Ge$_8$Sb$_2$Te$_{11}$ | 4 | 3.7 | 2.9 | 2.73 | 2.88 | 2.85 | [72] |
| AIMD; PBE, GTH (270) | Ge$_2$Sb$_2$Te$_5$ | 3.96 | 4.15 | 2.97 | 2.77 | 2.94 | 2.92 | [73] |
| AIMD; HSE03, GTH (270) | Ge$_2$Sb$_2$Te$_5$ | | | | 2.75/2.72[c] | 2.88/2.92[c] | 2.86 | [73] |
| AIMD; GGA, TM (168) | Ge$_1$Sb$_2$Te$_4$ | 3.5 | 3.8 | 2.9 | | | | [74] |
| AIMD; PBE, PAW (72) | Ge$_2$Sb$_2$Te$_5$ | 3.7 | 3.2 | 2.6 | 2.78 | 2.93 | | [75] |
| AIMD; BLYP, PAW (72) | Ge$_2$Sb$_2$Te$_5$ | 3.4 | 3 | 2.4 | 2.74 | 2.86 | | [75] |
| AIMD; HSE-q/h, PAW (72) | Ge$_2$Sb$_2$Te$_5$ | 3.7/3.7 | 3.2/3.1 | 2.4/2.2 | 2.67/2.62 | 2.86/2.79 | | [75] |
| AIMD; BLYP, GTH (459) | Ge$_2$Sb$_2$Te$_5$ | 3.74 | 4 | 2.84 | | | | [76] |
| AIMD; BLYP, TM (144) | Ge$_2$Sb$_2$Te$_5$ | 4.03 | 3.99 | 2.58 | 2.63 | 2.89 | | [77] |
| AIMD; PBE, TM (144) | Ge$_2$Sb$_2$Te$_5$ | 4.13 | 4.16 | 2.72 | 2.66 | 2.92 | | [77] |
| AIMD; BLYP, GTH (144) | Ge$_2$Sb$_2$Te$_5$ | 4.99 | 4.64 | 3.45 | 2.8 | 3.0 | | [77] |
| ML-MD; PBEsol, PAW (7200) | Ge$_2$Sb$_2$Te$_5$ | 4.5 | 3.7 | 3.1 | 2.8-2.82 (2.62-2.67[b]) | | | [78] |
| AIMD;PBEsol, TM, +vdW (300) | Ge$_6$Sb$_6$Te$_{88}$<br>Ge$_{10}$Sb$_{10}$Te$_{80}$<br>Ge$_{14}$Sb$_{14}$Te$_{72}$<br>Ge$_2$Sb$_2$Te$_5$ | 4.05<br>3.93<br>4.14<br>4.07 | 3.23<br>3.24<br>3.25<br>3.27 | 2.46<br>2.46<br>2.50<br>2.45 | <br><br><br>2.64 | <br><br><br>2.96 | <br><br><br>2.88 | [36] |
| AIMD; BLYP, TM (144/504) | Ge$_2$Sb$_2$Te$_5$ | 4.03/3.79 | 3.99/4.15 | 2.58/2.65 | 2.66/2.63 | 2.89 | -/2.95 | [79] |
| AIMD; BLYP, GTH, (144/504) | Ge$_2$Sb$_2$Te$_5$ | 4.99/3.84 | 4.64/3.45 | 3.45/2.63 | 2.8/2.74 | 3.0/2.89 | -/2.89 | [79] |

[a]This measurement cannot distinguish Te atoms from Sb atoms, thus Sb-Te and Te-Te distances were not resolved.

[b]Bond distance for tetrahedral Ge.

[c]At experimental/theoretical equilibrium density.

Figures

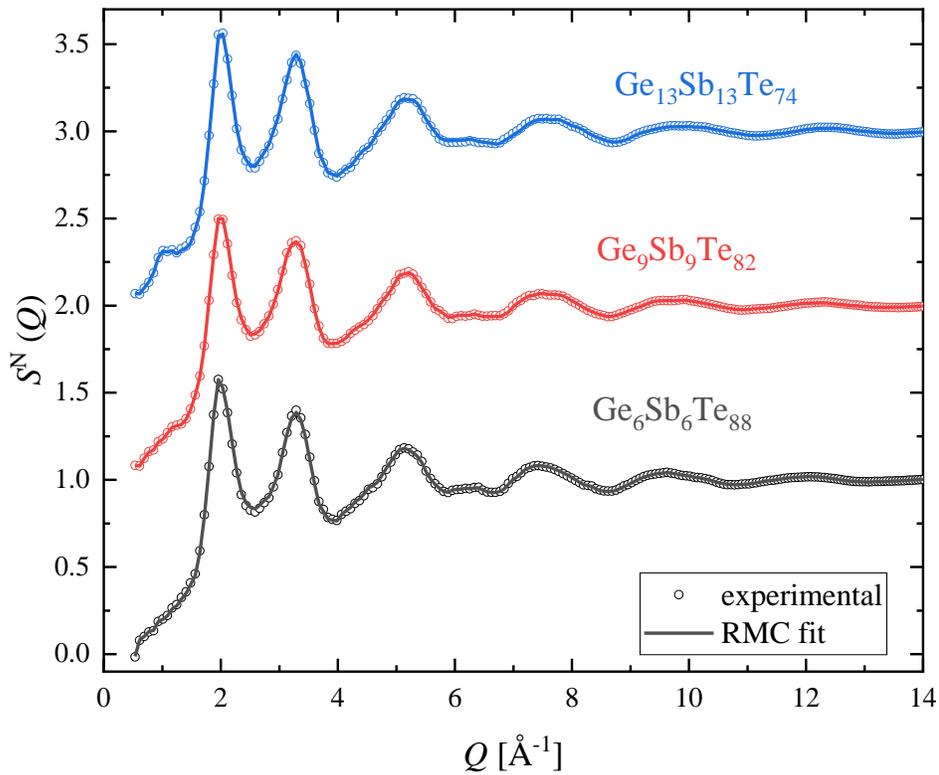

**Figure 1** ND structure factors (symbols) and fits (lines) of the final models obtained by RMC simulations of Ge-Sb-Te samples. (Curves of $Ge_9Sb_9Te_{82}$ and $Ge_{13}Sb_{13}Te_{74}$ glasses are shifted for clarity.)

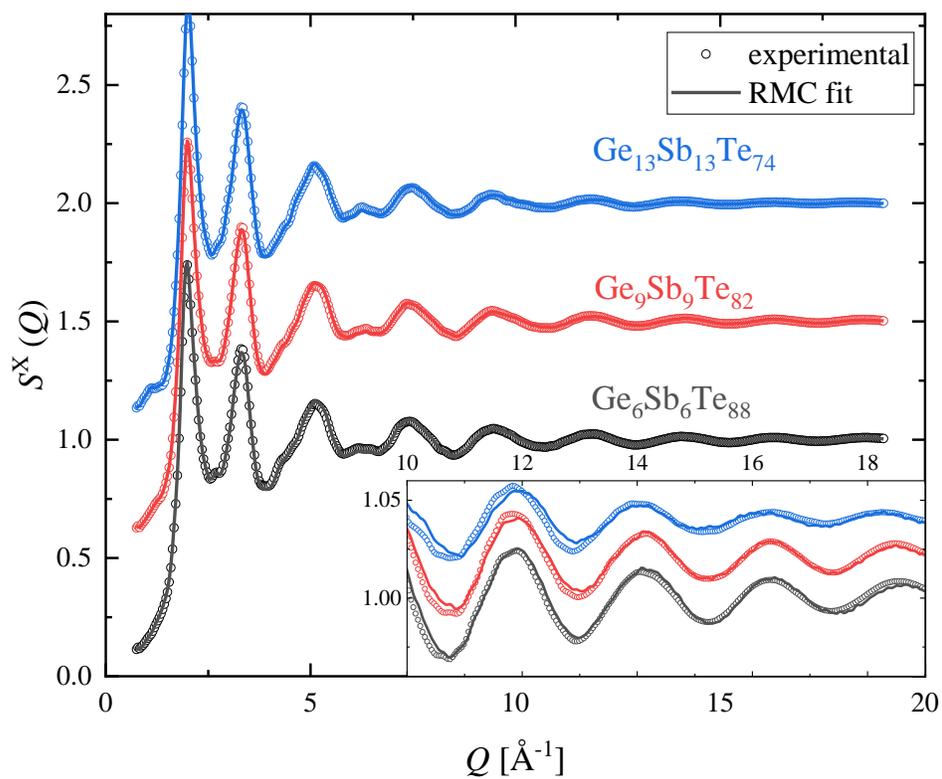

**Figure 2** XRD structure factors (symbols) and fits (lines) of the final models obtained by RMC simulations of Ge-Sb-Te samples. The inset is an enlargement of the curves at high $Q$ values. (The curves of $Ge_9Sb_9Te_{82}$ and $Ge_{13}Sb_{13}Te_{74}$ glasses are shifted for clarity.)

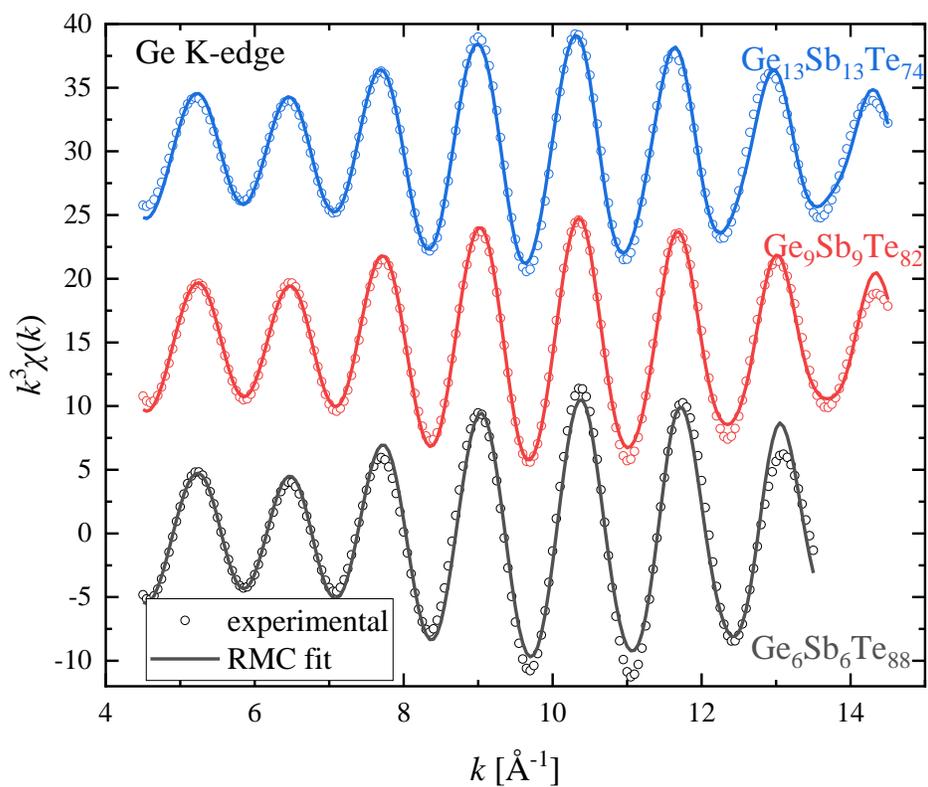

**Figure 3** $k^3$-weighted, filtered EXAFS spectra at Ge K-edge (symbols) and fits (lines) of the final models obtained by RMC simulations of Ge-Sb-Te samples. (The curves of $Ge_9Sb_9Te_{82}$ and $Ge_{13}Sb_{13}Te_{74}$ glasses are shifted for clarity.)

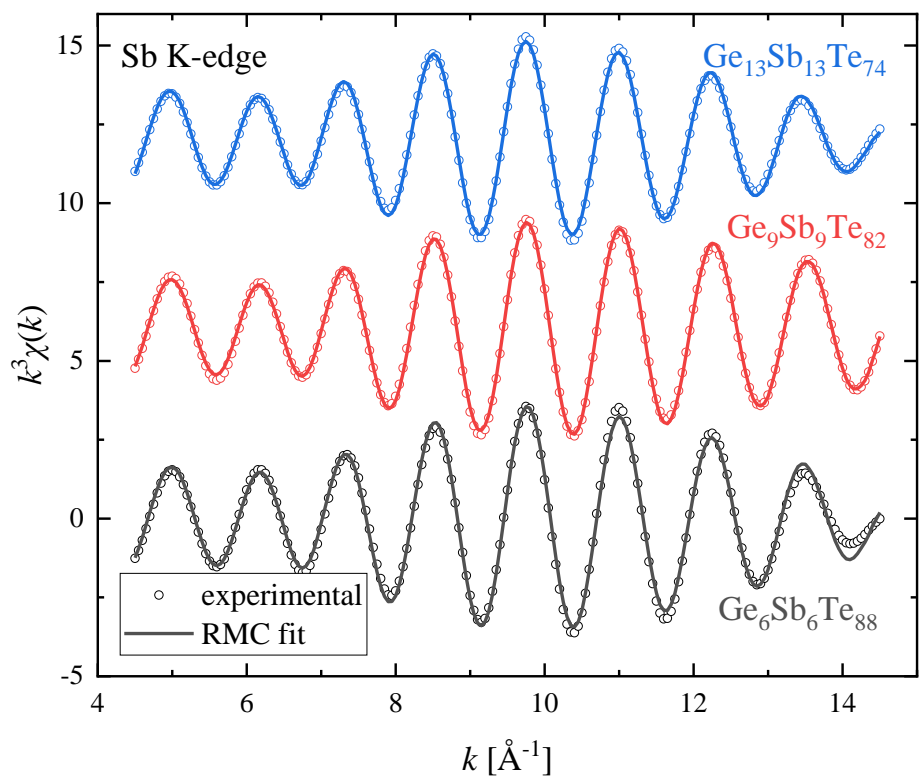

**Figure 4** $k^3$-weighted, filtered EXAFS spectra at Sb K-edge (symbols) and fits (lines) of the final models obtained by RMC simulations of Ge-Sb-Te samples. (The curves of Ge$_9$Sb$_9$Te$_{82}$ and Ge$_{13}$Sb$_{13}$Te$_{74}$ glasses are shifted for clarity.)

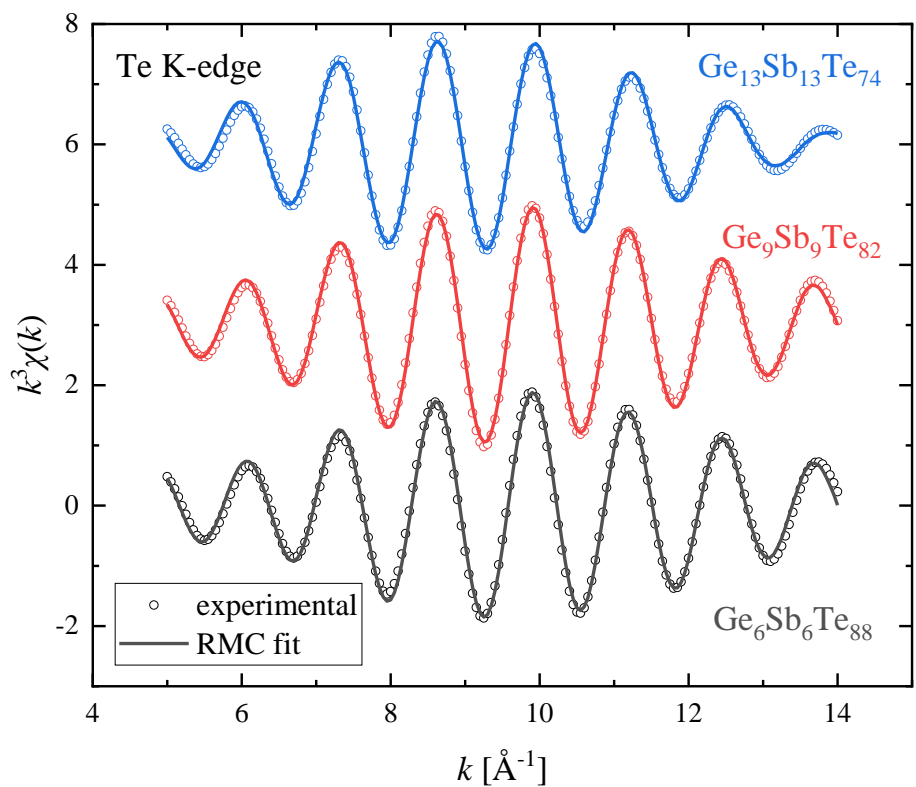

**Figure 5** $k^3$-weighted, filtered EXAFS spectra at Te K-edge (symbols) and fits (lines) of the final models obtained by RMC simulations of Ge-Sb-Te samples. (The curves of $Ge_9Sb_9Te_{82}$ and $Ge_{13}Sb_{13}Te_{74}$ glasses are shifted for clarity.)

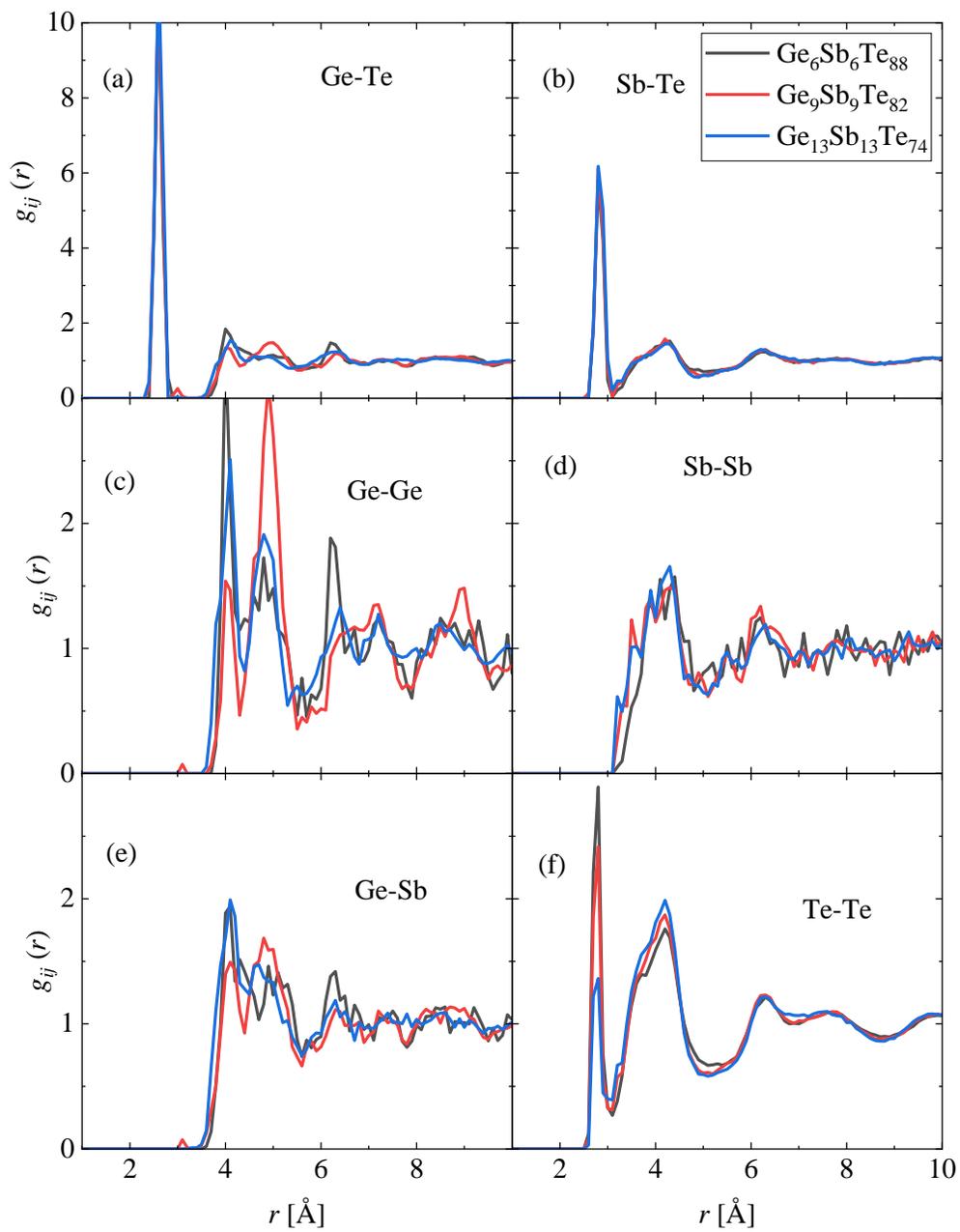

**Figure 6** Partial pair correlation functions of the $Ge_xSb_xTe_{100-2x}$ alloys obtained for the final model by RMC simulations.

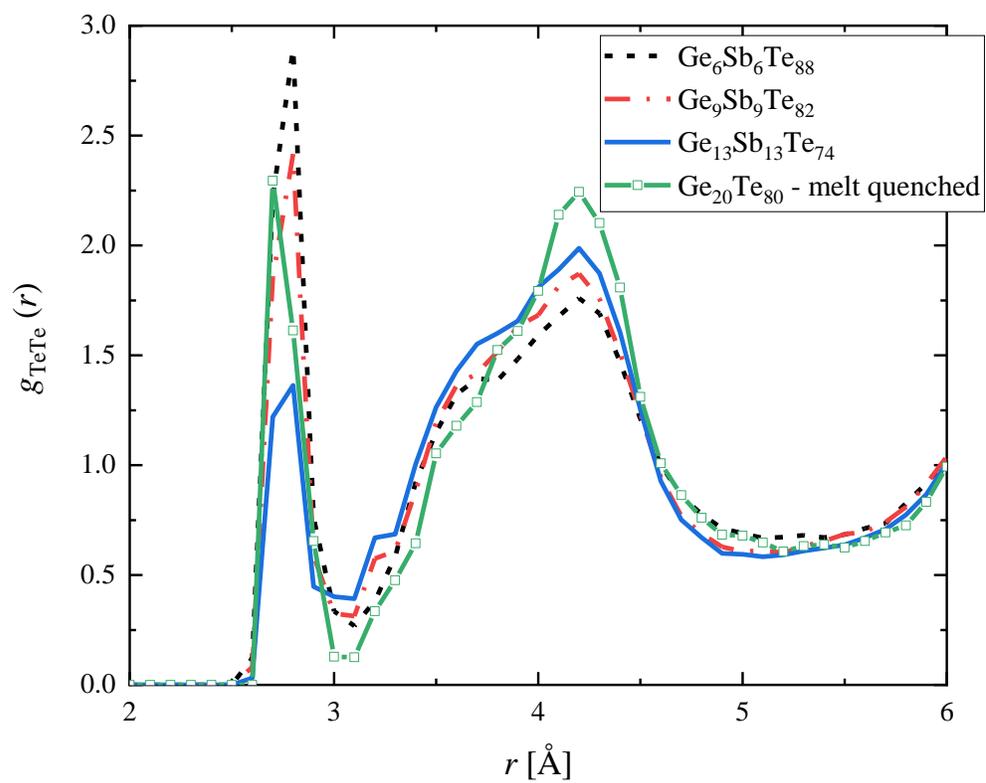

**Figure 7** Comparison of $g_{TeTe}(r)$ partial pair correlation functions of the $Ge_xSb_xTe_{100-2x}$ alloys investigated and those of amorphous $Ge_{20}Te_{80}$ obtained by melt quenching [8].

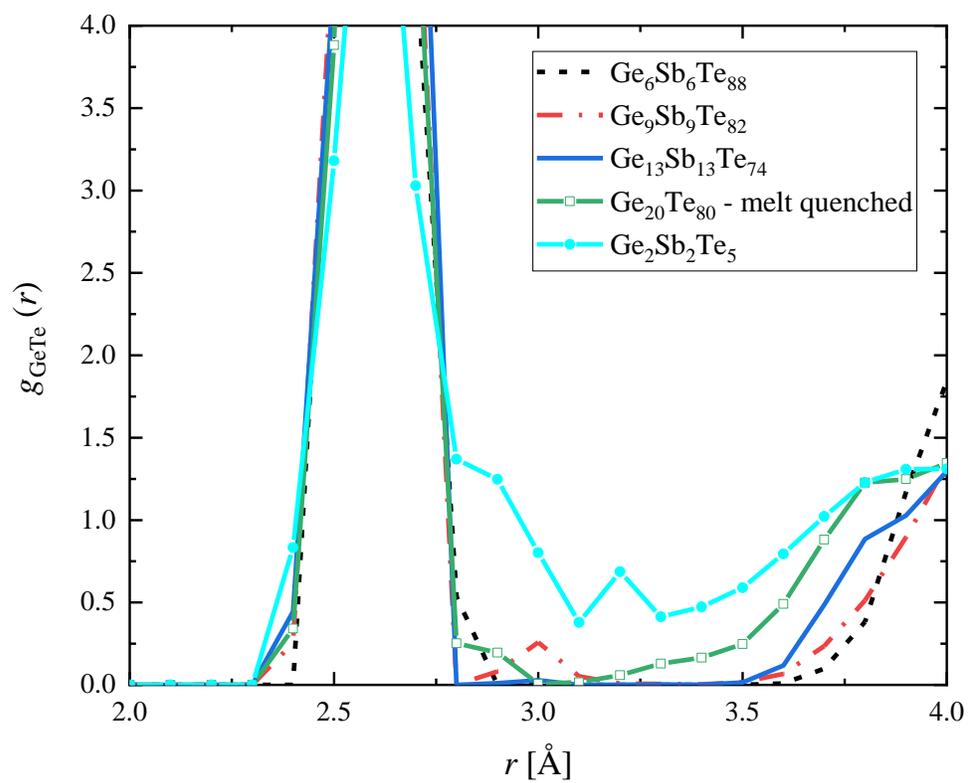

**Figure 8** Comparison of the $g_{GeTe}(r)$ partial pair correlation functions of the $Ge_xSb_xTe_{100-2x}$ alloys investigated with that of melt quenched binary $Ge_{20}Te_{80}$ [8] and sputtered amorphous $Ge_2Sb_2Te_5$ [12].

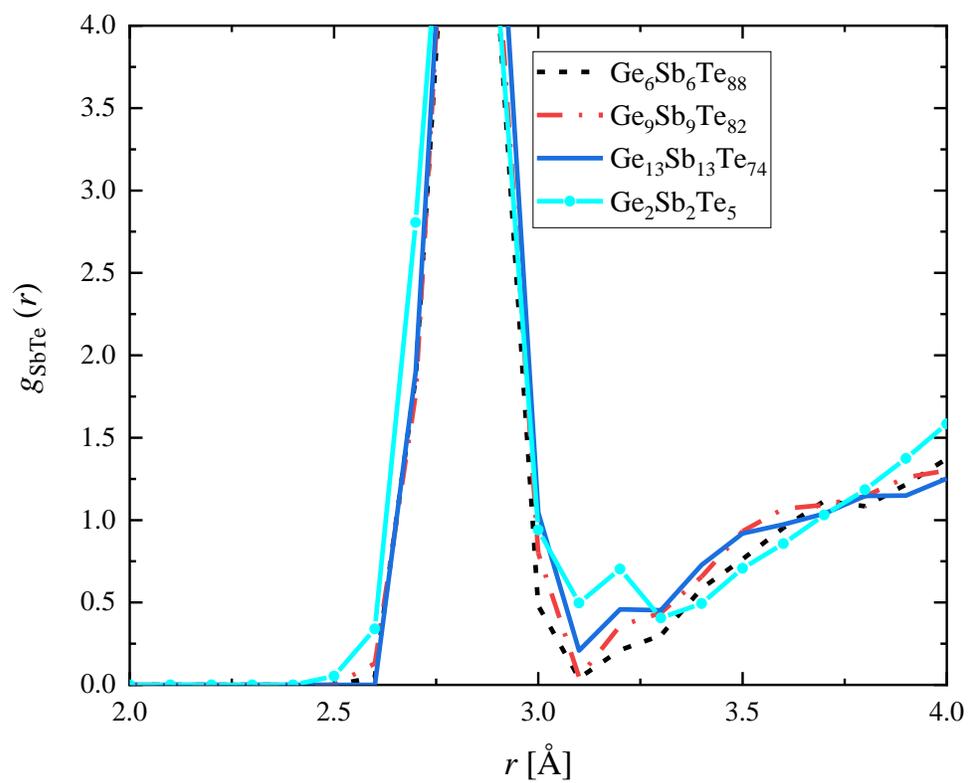

**Figure 9** Comparison of the $g_{SbTe}(r)$ partial pair correlation functions of evaporated Te-rich $Ge_xSb_xTe_{100-2x}$ alloys (present study) and that of sputtered amorphous $Ge_2Sb_2Te_5$ [12].

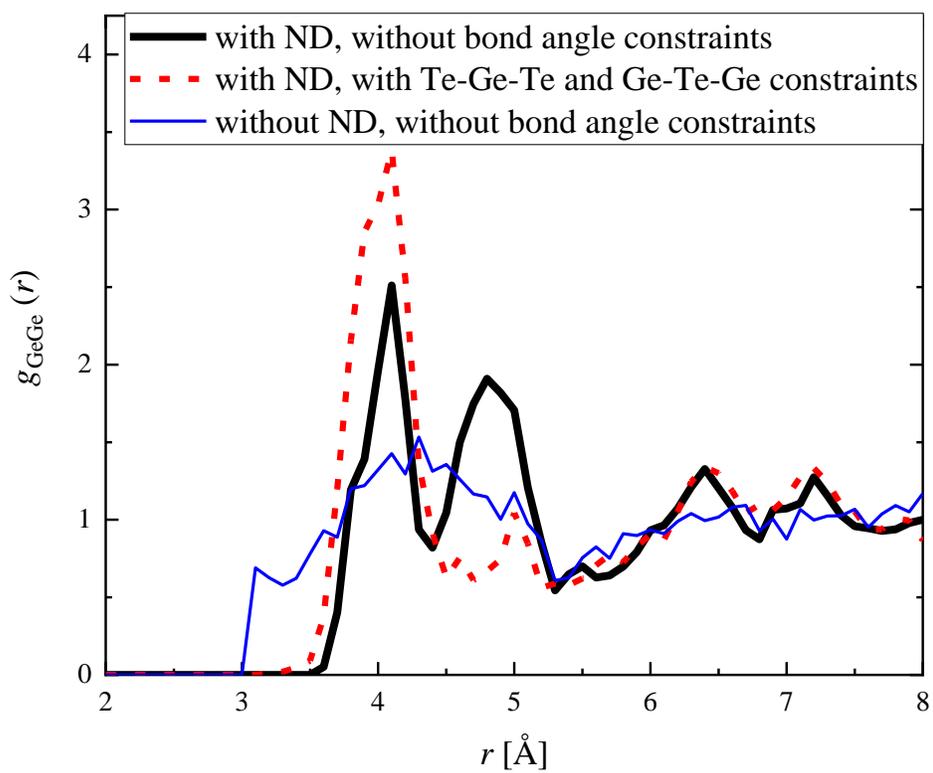

**Figure 10** The $g_{GeGe}(r)$ of $Ge_{13}Sb_{13}Te_{74}$ obtained with and without neutron diffraction data, and with using bond angle constraints. See the text for detailed description of bond angle constraints.

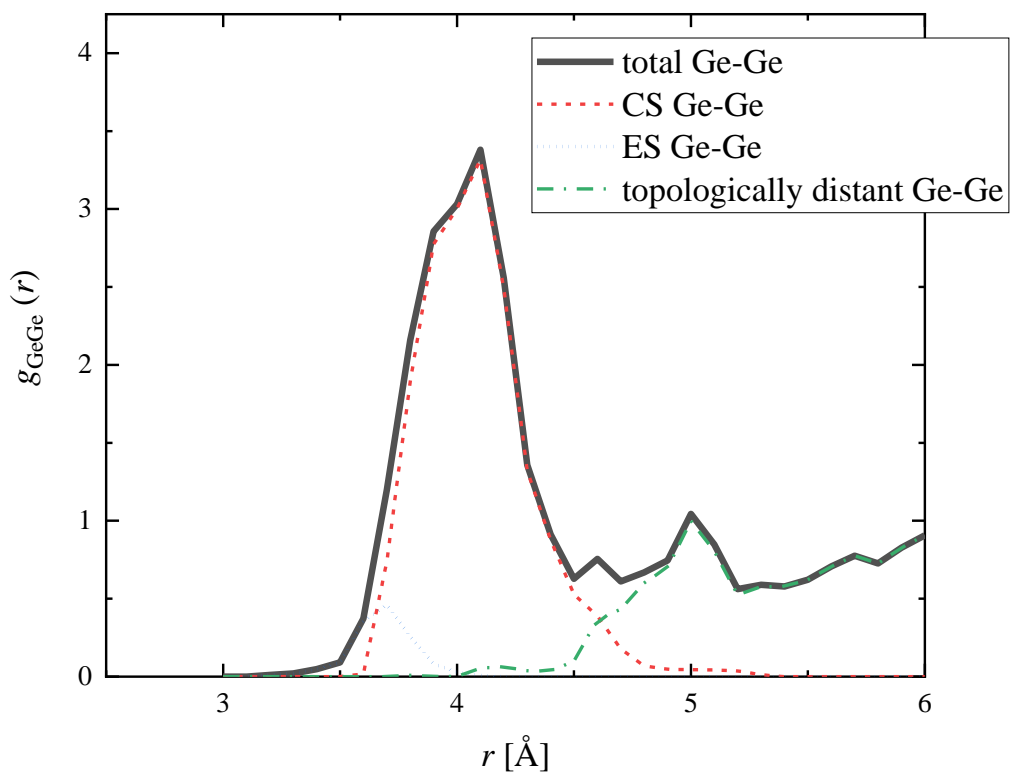

**Figure 11** Decomposition of the first peak of $g_{GeGe}(r)$ of $Ge_{13}Sb_{13}Te_{74}$ to contributions from corner-shared tetrahedra (CS), edge-shared tetrahedra (ES) and topologically distant Ge-Ge pairs.